\documentclass[12pt,a4paper]{article}%
\usepackage{amsfonts}
\usepackage{amsmath}
\usepackage{amssymb}
\usepackage{graphicx}%
\begin{document}

\title{Static Observers in Curved Spaces and Non-inertial Frames in Minkowski Spacetime}
\author{F. Dahia and P. J. Felix da Silva\\Departamento de F\'{\i}sica, Universidade Federal de Campina Grande,\\58109-970, PB, Brazil\\ }
\maketitle

\begin{abstract}
Static observers in curved spacetimes may interpret their proper
acceleration as the opposite of a local gravitational field (in
the Newtonian sense). Based on this interpretation and motivated
by the equivalence principle, we are led to investigate
congruences of timelike curves in Minkowski spacetime whose
acceleration field coincides with the acceleration field of static
observers of curved spaces. The congruences give rise to
non-inertial frames that are examined. Specifically we find, based
on the locality principle, the embedding of simultaneity
hypersurfaces adapted to the non-inertial frame in an explicit
form for arbitrary acceleration fields. We also determine, from
the Einstein equations, a covariant field equation that regulates
the behavior of the proper acceleration of static observers in
curved spacetimes. It corresponds to an exact relativistic version
of the Newtonian gravitational field equation. In the specific
case in which the level surfaces of the norm of the acceleration
field of the static observers are maximally symmetric
two-dimensional spaces, the energy-momentum tensor of the source
is analyzed.

\end{abstract}

\section{Introduction}

Based on the equivalence principle, it is expected that some physical features
of gravity can be mimicked by accelerated frames in Minkowski spacetime. The
Rindler frame, which is adapted to a family of uniformly accelerated
observers, is a famous example of a non-inertial system that simulates some
characteristics of a black hole's geometry\cite{misner,rindler}.

This frame has been widely investigated in the literature and here we are
going to start our discussion pointing out a peculiar aspect of the Rindler
frame. It is related to the remarkable characteristic that the proper
acceleration $a$ of Rindler observers, which is constant along their
worldlines, varies according to the law $a=1/\rho$ in relation to the
observers, where $\rho$ corresponds to the initial distance of the observer
with respect to the origin of an inertial frame. As it is well known, the
reason of this behavior is related to the geometric properties of the timelike
Killing field which generates the congruence.

On the other hand, from the physical point of view, it is very suggestive that
this particular dependence of $a$ and $\rho$ is connected to the behavior of
static observers in Schwarzschild geometry in the vicinity of the horizon.
Indeed, if $\rho$ denotes the radial distance of an observer to the horizon,
then, the proper acceleration the observers need in order to stay at rest in
their position close to the horizon is proportional to $1/\rho$. Therefore the
Rindler congruence and the static Schwarzschild observers have the same
acceleration field $a(\rho)$. This equivalence reinforce the linkage between
Rindler frame and the exterior of a black hole.

However, it happens that the inverse law holds only in the region where $\rho$
is small in comparison to the Schwarzschild radius of the black hole. Indeed,
static observers at long distances are submitted to an acceleration field
proportional to $1/\rho^{2}$, (i.e, they obey the inverse square law of
Newtonian gravitation) and, in the intermediate domain, the dependence of $a$
in terms of $\rho$ is described by a much more complicated function.

We can say, then, that static observers in the Schwarzschild geometry are
accelerated according to a cumbersome function $a(\rho)$ which reduces to the
power law $1/\rho$ only in the region near the horizon. All these
considerations led us to turn our attention to the study of a congruence of
timelike curves whose acceleration field $a\left(  \rho\right)  $ coincides
with the acceleration field of static observers in the Schwarzschild
spacetime. In this case, only in the range of small values of $\rho$, the
congruence defined in Minkowski spacetime will be equivalent to the worldlines
of Rindler observers.

Of course this discussion is not limited to the Schwarzschild spacetime and
can be extend to encompass any static spacetime. In this paper, motivated by
these ideas, we are going to investigate congruences in Minkowski spacetime
that obey a generic law $a(\rho)$, focusing on the problem of determining a
coordinate system adapted to the non-inertial frame associated to these congruences.

A fundamental ingredient in this formalism is the determination of the
simultaneity hypersurfaces relative to the non-inertial frame. As we shall
see, the embedding functions of the simultaneity hypersurfaces into Minkowski
spacetime are explicitly obtained for any arbitrary function $a\left(
\rho\right)  $, by using the locality principle \cite{mashhoon,longhi}. The
formulation of this general scheme allow us to discuss the correspondence
between non-inertial frames in Minkowski spacetime and static observers of any
static spacetimes. The particular case concerning the Schwarzschild spacetime,
due to its relevance, is investigated in detail.

Considering the importance of the acceleration of static observers in this
context, it is desirable to find a field equation that regulates the behavior
of $a$. By using the embedding formalism we obtain, from the Einstein
equations, a covariant field equation for the proper acceleration that has a
clear physical interpretation. Indeed, from the perspective of static
observers, non-interacting bodies are accelerated as if they were under the
influence of a gravitational field whose intensity is precisely equal to $a$
but in the opposite direction. Thus, the field equation for $a$ may be
understood as that the exact relativistic version of the gravitational field
equation of Newtonian theory, as we shall see.

By admitting additional symmetries, we can focus our attention in a special
class of static spacetimes characterized by the fact that the level surfaces
of the norm of the proper acceleration field are maximally symmetric
two-dimensional spaces. This assumption simplifies greatly the Einstein
equations and as a consequence the field equation can be totally written in
terms of the extrinsic and intrinsic curvature of the level surfaces and the
acceleration field.

The general form of the energy-momentum tensor of the possible sources of this
special kind of static spacetimes can be deduced from the Einstein equation.
As we shall see, it corresponds to non-viscous fluids that may be anisotropic.
Indeed, the pressure in the parallel direction $\left(  p_{\Vert}\right)  $ to
the acceleration field may be different from the pressure the fluid exerts in
the orthogonal direction $\left(  p_{\bot}\right)  $. In the particular case
in which $p_{\bot}=0$ and the parallel pressure satisfies the state equation
characteristic of a false vacuum state, the produced spacetime possesses an
intriguing property: the static observers are accelerated precisely according
to the Rindler law, although they are in a curved spacetime.

This paper is organized as follows. In the first section we review the Rindler
frame in order to establish the notation and the techniques necessary for the
construction of a coordinate system adapted to a non-inertial frame. In the
second section, we study the induced geometry in the simultaneity
hypersurfaces relative to the accelerated congruence. Next, in the third
section, the idea of establishing a mapping between non-inertial frame and
static observers is elaborated and the particular case related to the
Schwarzschild spacetime is discussed in detail. The fourth section is
dedicated to determine the field equation that regulates the acceleration
field of static observers in curved spacetimes and to study the
energy-momentum tensor of the sources that generate a special class of static spacetimes.

\section{Accelerated Observers in Minkowski spacetime}

For the sake of simplicity, let us initially consider a two-dimensional
Minkowski spacetime mapped by coordinates $(t,x)$ of an inertial system $S.$
Consider now an observer submitted to a constant proper acceleration $a(\rho
)$, where $\rho$ corresponds to the initial distance of the observer with
respect to the origin of $S.$ The worldline of this observer, that we will
denote by $O_{\rho}$ henceforth, is described by the following parametric
curve:%
\begin{align}
t\left(  \tau_{\rho}\right)   &  =\frac{1}{a\left(  \rho\right)  }\sinh\left(
a(\rho)\tau_{\rho}\right) \label{eqt}\\
x\left(  \tau_{\rho}\right)   &  =\frac{1}{a\left(  \rho\right)  }\cosh\left(
a\left(  \rho\right)  \tau_{\rho}\right)  -\frac{1}{a\left(  \rho\right)
}+\rho\label{eqx}%
\end{align}
where $\tau_{\rho}$ is the proper time of $O_{\rho}$. It can be directly
checked that the proper acceleration of the observer is indeed $a\left(
\rho\right)  $ and that in the initial position, $x(0)=\rho$ , the observer is
instantaneously at rest.

In the spacetime diagram of $S,$ this worldline is a hyperbola given by the
equation%
\begin{equation}
\left(  x-b\left(  \rho\right)  \right)  ^{2}-t^{2}=\frac{1}{a\left(
\rho\right)  } \label{hyperbola}%
\end{equation}
where $a^{-1}\left(  \rho\right)  $ is the distance from the vertex of the
hyperbola to its center $b\left(  \rho\right)  =\rho-\frac{1}{a\left(
\rho\right)  }$ , located at the $x$-axis.

The motion equations (\ref{eqt}) and (\ref{eqx}) may also be understood
according to a different perspective that is more appropriate to our
discussion. They can be considered as a set of equations which describe a
family of accelerated observers, where for each value of $\rho$ corresponds a
different observer. Thus, with respect to the congruence, the coordinate
$\rho$ can be viewed as a parameter that identify each member of the family.
It is important to emphasize that each observer suffers an uniform
acceleration along its worldline. However, distinct observers are submitted to
different accelerations according to a generic law $a\left(  \rho\right)  $.

We want now to construct a coordinate system adapted to this non-inertial
frame. A crucial element in this discussion is the notion of simultaneity
associated to the frame. There is no privileged or natural way to establish it
and as a consequence there are a multitude of options that can be found in the
literature. Every possible way leads to a different kind of coordinate
system\cite{marzlin,letaw,gao,pauri,alba,alba1,alba2,bini,bini2,minguzzi,minguzzi2,minguzzi3,minguzzi4,huang,rey,patricio}%
.

Here we will follow a very reasonable idea known as the locality principle.
According to this principle the accelerated observer is instantaneously
equivalent to the co-moving inertial observer. The reason is that they share
the same position and velocity instantaneously and therefore, as both
observers have the same physical state according to non-quantum mechanics,
they should be considered physically equivalent at that moment.

If we admit this hypothesis as being valid then it is natural to assume that
the set of simultaneous events for the accelerated observer coincides locally
with the set of simultaneous events relative to the co-moving inertial observer.

In Minkowski spacetime this set could be identified as follows. Consider
$S^{\prime}\left(  \rho,\tau\right)  $ the co-moving inertial frame relative
to the observer $O_{\rho}$ at a certain instant of time $\tau$. Obviously (by
definition), the accelerated observer will be find in $O^{\prime}$, the origin
of $S^{\prime}\left(  \rho,\tau\right)  $, with null velocity instantaneously.
Now, with respect to the inertial frame $S$, let $R^{\mu}$ be the components
of the position vector of a generic event $E$ and $R_{0}^{\mu}$ the vector
that localize the origin of $S^{\prime}\left(  \rho,\tau\right)  $ - which
corresponds also to the spacetime position of $O_{\rho}$ at instant $\tau$.
The event $E$ will be considered simultaneous to $O^{\prime}$, in the inertial
frame $S^{\prime}\left(  \rho,\tau\right)  $, if the relative position of $E$
(which is given by $\Delta R^{\mu}=R^{\mu}-R_{0}^{\mu}$) has no timelike
component when decomposed in the coordinate basis associated to $S^{^{\prime}%
}$, since, in this case, there will be no timelike separation between the
events $E$ and $O^{\prime}$ as viewed from $S^{^{\prime}}$ frame.

Considering that the direction of the timelike axis of the frame $S^{^{\prime
}}$ coincides with the direction of the vector $U^{\mu}$ (the proper velocity
of $O_{\rho}$), then, it follows that the condition of simultaneity is
equivalent to the orthogonality condition between the relative position and
$U^{\mu}$ in the Minkowski metric :%
\begin{equation}
U_{\mu}\left(  R^{\mu}-R_{0}^{\mu}\right)  =0 \label{simeq}%
\end{equation}
In the two-dimensional case, the solution of the above equation, obviously,
corresponds exactly to the $x^{\prime}-$axis of $S^{\prime}.$

When we are dealing with an observer $O_{\rho}$ of the accelerated congruence,
then, based on the principle of locality, we should be aware that the validity
of equation (\ref{simeq}) is local, since $U^{\mu}$ changes in time (the
observer is accelerated) and with respect to the space (different observers
are submitted to different accelerations). So the equation that defines
simultaneous events relative to accelerated frame is an infinitesimal version
of the above equation, i.e.,%
\begin{equation}
U_{\mu}dR^{\mu}=0 \label{simdif}%
\end{equation}
This means that the proper velocity of each observer must be orthogonal to any
infinitesimal displacement in the simultaneity section, $dR^{\mu}$, in every
point of the hypersurface. In other words, proper velocity is the normal
vector of the simultaneity sections.

Let us assume that the simultaneity sections can be described by a function
$\varphi$ defined in Minkowski spacetime. More precisely, let us admit that
each equation $\varphi=$\textit{constant} determines a hypersurface in
spacetime which corresponds to a simultaneity section adapted to the
accelerated observers. We know that the normal vector of the hypersurface is
proportional to the gradient of the function $\varphi$. Therefore, in order to
satisfy equation (\ref{simdif}), we must have:%
\begin{equation}
\frac{1}{\left\vert \nabla\varphi\right\vert }\frac{\partial\varphi}{\partial
x^{\mu}}=U_{\mu} \label{grad}%
\end{equation}
where $\left\vert \nabla\varphi\right\vert $ is the modulus of the norm of the
gradient in Minkowski metric.

For the sake of simplicity let us use coordinates $\rho$ and $\tau$ to
localize events\footnote{Henceforth we will write $\tau$ in the place of
$\tau_{\rho}$ in order to make the notation simpler.}. Of course, whenever it
is necessary, we can, from equations (\ref{eqt}) and (\ref{eqx}), obtain the
corresponding $(t,x)$-coordinates in the frame $S$.

Thus, assuming we have the function $\varphi\left(  \rho,\tau\right)  $,
equation (\ref{grad}) can be written, by using a well-known relation involving
partial derivatives, in the following way%
\begin{equation}
-\left(  \frac{\partial\tau}{\partial\rho}\right)  _{\varphi}=\frac{\left(
\frac{\partial\varphi}{\partial\rho}\right)  _{\tau}}{\left(  \frac
{\partial\varphi}{\partial\tau}\right)  _{\rho}}=\frac{U_{\rho}}{U_{\tau}}%
\end{equation}
On the other hand, by using equations (\ref{eqt}) and (\ref{eqx}) as
transformation equations, $U_{\rho}$ and $U_{\tau}$ can be determined from the
components of the proper velocity $U_{t}=-\cosh a\tau$ and $U_{x}=\sinh a\tau
$. We find
\begin{align}
U_{\tau}  &  =\left(  \frac{\partial t}{\partial\tau}\right)  _{\rho}%
U_{t}+\left(  \frac{\partial x}{\partial\tau}\right)  _{\rho}U_{x}=-1\\
U_{\rho}  &  =\left(  \frac{\partial t}{\partial\rho}\right)  _{\tau}%
U_{t}+\left(  \frac{\partial x}{\partial\rho}\right)  _{\tau}U_{x}%
=-\frac{a^{\prime}\tau}{a}+\left(  \frac{a^{\prime}}{a^{2}}+1\right)
\sinh\left(  a\tau\right)
\end{align}
Therefore, the equation for simultaneity hypersurface takes the form:%
\begin{equation}
\left(  \frac{\partial\tau}{\partial\rho}\right)  _{\varphi}=-\frac{a^{\prime
}\tau}{a}+\left(  \frac{a^{\prime}}{a^{2}}+1\right)  \sinh\left(
a\tau\right)
\end{equation}

In order to solve this equation it is convenient to introduce a new
coordinate
\begin{equation}
\eta=a\left(  \rho\right)  \tau
\end{equation}
in terms of which the above equation reduces to%
\begin{equation}
\left(  \frac{\partial\eta}{\partial\rho}\right)  _{\varphi}=\left(
\frac{a^{\prime}}{a}+a\right)  \sinh\left(  \eta\right)
\end{equation}

This equation can be directly integrated and the solution is
\begin{equation}
\tau=\frac{1}{a}\ln\left[  \frac{1+f\left(  \varphi\right)  a\exp\left(  \int
ad\rho\right)  }{1-f\left(  \varphi\right)  a\exp\left(  \int ad\rho\right)
}\right]  \label{sim section}%
\end{equation}
where $f\left(  \varphi\right)  $ is some arbitrary function. For each
specified value $\varphi=const$, equation (\ref{sim section}) gives us the
coordinates $\left(  \tau,\rho\right)  $ of the simultaneous events relative
to the non-inertial frame.

The simultaneity hypersurfaces can also be described by parametric equations
in the coordinate system of the inertial frame $S$. Indeed, by using equations
(\ref{eqt}), (\ref{eqx}) and (\ref{sim section}), we find:%

\begin{align}
t  &  =F_{\varphi}\left(  \rho\right)  \equiv\frac{2}{a}\left[  \frac{f\left(
\varphi\right)  a\exp\left(  \int ad\rho\right)  }{1-\left(  f\left(
\varphi\right)  a\exp\left(  \int ad\rho\right)  \right)  ^{2}}\right]
\label{tsim}\\
x  &  =G_{\varphi}\left(  \rho\right)  \equiv\frac{2}{a}\left[  \frac{\left(
f\left(  \varphi\right)  a\exp\left(  \int ad\rho\right)  \right)  ^{2}%
}{1-\left(  f\left(  \varphi\right)  a\exp\left(  \int ad\rho\right)  \right)
^{2}}\right]  +\rho\label{xsim}%
\end{align}

As we have mentioned, the coordinate $\varphi$ can be used to label the
simultaneity hypersurfaces. For each value $\varphi=const$ corresponds a
simultaneity hypersurface and the equations above give the embedding functions
of the hypersurface in the Minkowski spacetime. By varying the coordinate
$\rho$, we can find the image of the hypersurfaces embedded in the spacetime
diagram of $S$ frame.

It is also possible to connect $\varphi$ with the proper time that is measured
by observers. Choosing a certain particular observer to serve as a reference,
let us say $\rho_{0},$ then, from equation (\ref{sim section}), we find the
following relation%
\begin{equation}
f\left(  \varphi\right)  =\frac{\tanh(\frac{a\left(  \rho_{0}\right)  \tau
_{0}}{2})}{a\left(  \rho_{0}\right)  \exp\left(  \int^{\rho_{0}}ad\rho\right)
}%
\end{equation}
Thus, the simultaneity section may be labelled, in an equivalent way, by the
proper time $\tau_{0}$ of the observer $\rho_{0}$.

The equations (\ref{tsim}) and (\ref{xsim}) are very general in the sense that
they are applicable to any function $a\left(  \rho\right)  $. In the appendix,
we analyze several simple cases of physical interest. Figure (1) shows that
the general formulas (\ref{tsim}) and (\ref{xsim}) reproduce the Rindler
congruence when we take $a=1/\rho$. In the second figure, we have considered a
congruence that follows the inverse square law. And at last, the uniform
acceleration field $a=cte$, which was also considered with a different notion
of simultaneity in the literature \cite{huang,huang1}, is examined in Figure (3).

\subsection{Geometry of the simultaneity hypersurfaces}

A generalization to (3+1)-dimensions can be immediately obtained as soon as we
have established the symmetry of the spatial distribution of the observers.
This is directly connected with the interpretation of $\rho$ as a spatial
coordinate. For instance, if the spherical symmetry is admitted, then, it is
implicitly assumed that the observers are performing a radial motion and
consequently $\rho$ plays the role of a radial coordinate. It follows then
that the simultaneity hypersurfaces are described by the equations:
\begin{align}
t  &  =F_{\varphi}\left(  \rho\right) \label{st}\\
x  &  =G_{\varphi}\left(  \rho\right)  \sin\theta\cos\phi\label{sx}\\
y  &  =G_{\varphi}\left(  \rho\right)  \sin\theta\sin\phi\label{sy}\\
z  &  =G_{\varphi}\left(  \rho\right)  \cos\theta\label{sz}%
\end{align}
However, in the case of cylindrical symmetry, the embedding functions are%
\begin{align}
t  &  =F_{\varphi}\left(  \rho\right) \label{ct}\\
x  &  =G_{\varphi}\left(  \rho\right)  \cos\phi\label{cx}\\
y  &  =G_{\varphi}\left(  \rho\right)  \sin\phi\label{cy}\\
z  &  =z^{\prime} \label{cz}%
\end{align}
while the plane symmetry leads to the following embedding map%
\begin{align}
t  &  =F_{\varphi}\left(  \rho\right) \label{pt}\\
x  &  =G_{\varphi}\left(  \rho\right) \label{px}\\
y  &  =y^{\prime}\label{py}\\
z  &  =z^{\prime} \label{pz}%
\end{align}

At this point, once we have already described the hypersurface by the
embedding functions, we are able now to study the induced geometry in this
hypersurface. With this purpose, let us first recall some elements of the
embedding formalism. Let $\psi:\Sigma\rightarrow M$ be an embedding map of a
hypersurface $\Sigma$ in a manifold $M$. Given a coordinate system $\left\{
\xi^{i}\right\}  $ on $\Sigma$ and $\left\{  x^{\alpha}\right\}  $ on $M$, the
embedding functions can be explicitly written as%
\begin{equation}
x^{\alpha}=\psi^{\alpha}\left(  \xi^{1},\xi^{2},\xi^{3}\right)
\end{equation}

The differential $df$, which is injective by definition, maps vectors of the
tangent space of $\Sigma$ into vectors of the tangent space of $M.$ In
particular, the basis vectors can be written as
\begin{equation}
\frac{\partial}{\partial\xi^{a}}=e_{a}^{\alpha}\frac{\partial}{\partial
x^{\alpha}}%
\end{equation}
where
\begin{equation}
e_{a}^{\alpha}=\frac{\partial\psi^{\alpha}}{\partial\xi^{a}} \label{e}%
\end{equation}

If the manifold $M$ is equipped with a metric, $g_{\alpha\beta}$, then, the
embedding induces a metric $h_{ab}$ in the hypersurface, which, in terms of
the coordinate basis, is given by%
\begin{equation}
h_{ab}=e_{a}^{\alpha}e_{b}^{\beta}g_{\alpha\beta} \label{hab}%
\end{equation}

The normal vector of the hypersurface $\Sigma$ relative to $M$ is, in the case
of our congruence, the timelike vector $U^{\alpha}.$ The orthogonality
condition with respect to the tangent vector $\frac{\partial}{\partial\xi^{a}%
}$ can be expressed as $e_{a}^{\alpha}U_{\alpha}=0$.

In the present formalism an important concept is that of a projection tensor
$\Pi_{\alpha\beta}$, which maps vectors of the tangent space of $M$ onto the
tangent space of the hypersurface $\Sigma.$ Its components are defined as%
\begin{equation}
\Pi_{\alpha\beta}=g_{\alpha\beta}+U_{\alpha}U_{\beta} \label{pi}%
\end{equation}

It is clear that, for any vector $V^{\alpha}$, the projection $\Pi_{\beta
}^{\alpha}V^{\beta}$ can be considered as a vector of the tangent space of
$\Sigma$. Therefore it can be written in terms of the basis $\left\{
\frac{\partial}{\partial\xi^{a}}\right\}  $. In particular, the projection of
$\partial_{\alpha}$ can be written as $e_{\alpha}^{a}\partial_{a}$, for some
'vielbein' $e_{\alpha}^{a}$. Thus, we have%
\begin{equation}
\Pi_{\alpha}^{\gamma}\frac{\partial}{\partial x^{\gamma}}=e_{\alpha}^{a}%
\frac{\partial}{\partial\xi^{a}} \label{DA}%
\end{equation}

Considering this relation and taking the inner product with vectors of the
basis $\left\{  \partial_{\beta}\right\}  $, it is possible to show that
$e_{\alpha}^{a}$ is related to $e_{a}^{\alpha}$ according to the following formula:%

\begin{equation}
e_{\alpha}^{a}=g_{\alpha\beta}h^{ab}e_{b}^{\beta}%
\end{equation}

The embedding also induces a covariant derivative in $\Sigma.$ If $\nabla$ is
the covariant derivative defined in $M$, compatible with $g_{\alpha\beta},$
then, by using the projection tensor, we can define a covariant derivative in
$\Sigma$ in the following way
\begin{equation}
^{\left(  3\right)  }\nabla_{\beta}v^{\alpha}=e_{A}^{\alpha}e_{\beta}%
^{B}\nabla_{B}v^{A} \label{CovD}%
\end{equation}
where $v^{\alpha}$ is some vector which belongs to the tangent space of
$\Sigma$ (more precisely, an extension of the vector). It can be checked that
the induced covariant derivative is also compatible with the induced metric
$h_{ab}$ \cite{manfredo}.

Another important concept in the context of the embedding formalism is that of
extrinsic curvature $K_{ab}$. Roughly speaking, we can say that it measures
the variation of the normal vector along tangent directions of the
hypersurface $\Sigma$. In terms of its components, we have the following
definition:%
\begin{equation}
K_{ab}=e_{a}^{\alpha}e_{b}^{\beta}\nabla_{\beta}U_{\alpha} \label{K}%
\end{equation}

From the induced covariant derivative, the intrinsic Riemann tensor $^{\left(
3\right)  }R_{abcd}$ of the hypersurface $\Sigma$ can be naturally
constructed. As it is well known, from relation (\ref{CovD}), the tensor
$^{\left(  3\right)  }R_{abcd}$ can be put in connection with the components
of the Riemnan tensor $R_{\alpha\beta\gamma\delta}$ defined in $M$, according
to the Gauss equation \cite{manfredo}:%
\begin{equation}
^{(3)}R_{abcd}=e_{a}^{\alpha}e_{b}^{\beta}e_{c}^{\gamma}e_{d}^{\delta
}R_{\alpha\beta\gamma\delta}+\left(  K_{ad}K_{bc}-K_{ac}K_{bd}\right)
\end{equation}
There is also the Codazzi equation that gives the variation of
extrinsic
curvature on the hypersurface $\Sigma$ \cite{manfredo}:%
\begin{equation}
^{\left(  3\right)  }\nabla_{c}K_{ab}-{}^{\left(  3\right)  }\nabla_{b}%
K_{ac}=U^{\mu}e_{a}^{\alpha}e_{b}^{\beta}e_{c}^{\gamma}R_{\mu\alpha\beta
\gamma}%
\end{equation}

In our case the ambient space $M$ is the Minkowski space, so these equations
reduce to:%
\begin{align}
^{\left(  3\right)  }R_{abcd}  &  =K_{ad}K_{bc}-K_{ac}K_{bd}\\
^{\left(  3\right)  }\nabla_{c}K_{ab}  &  ={}^{\left(  3\right)  }\nabla
_{b}K_{ac}%
\end{align}

After we have briefly reviewed these general concepts regarding
the embedding formalism, let us now turn our attention to our
particular embedding maps in order to study the induced geometry
in $\Sigma$. By
using the embedding functions, given by the sets of equations (\ref{st}%
)-(\ref{sz}), (\ref{ct})-(\ref{cz}) and (\ref{pt})-(\ref{pz}), and identifying
explicitly the intrinsic coordinates for each embedding map, we can directly
calculate $e_{a}^{\alpha}$ and $K_{ab}$.

First we are going to consider the case of spherical symmetry. Of course, the
intrinsic coordinates of the simultaneity hypersurface are: $\xi^{1}=\rho
,\xi^{2}=\theta,\xi^{3}=\phi$. In this coordinate system the induced metric in
$\Sigma$, according to the equation (\ref{hab}), is given by%
\begin{equation}
dl^{2}=\left[  \frac{a^{\prime}}{a^{2}}-\left(  1+\frac{a^{\prime}}{a^{2}%
}\right)  \cosh\left(  \eta\right)  \right]  ^{2}d\rho^{2}+G_{\varphi}%
^{2}\left(  \rho\right)  \left[  d\theta^{2}+\sin^{2}\theta d\phi^{2}\right]
\label{dls}%
\end{equation}
where $\eta=a\tau$ is given by equation (\ref{sim section}).

We can also verify that the extrinsic curvature is diagonal and its non-null
components are the following%
\begin{align}
K_{\rho\rho}  &  =\left(  \frac{a^{\prime}}{a}+a\right)  \left[
\frac{a^{\prime}}{a^{2}}-\left(  1+\frac{a^{\prime}}{a^{2}}\right)  \cosh
\eta\right]  \sinh\left(  \eta\right) \label{sK1}\\
K_{\theta\theta}  &  =\left[  \frac{1}{a}\left(  \cosh\eta-1\right)
+\rho\right]  \sinh\eta\label{sK2}\\
K_{\phi\phi}  &  =\left(  \sin^{2}\theta\right)  K_{\theta\theta} \label{sK3}%
\end{align}

As it is well known, in a 3-dimensional space, the Riemann tensor has only 6
algebraically independent components. In this present case, all non-null
components of $^{\left(  3\right)  }R_{abcd}$ can be determined by this
following set of components:%
\begin{align}
^{\left(  3\right)  }R_{\rho\theta\rho\theta}  &  =-K_{\rho\rho}%
K_{\theta\theta}\label{sRiem1}\\
^{\left(  3\right)  }R_{\rho\phi\rho\phi}  &  =-K_{\rho\rho}K_{\theta\theta
}\left(  \sin^{2}\theta\right) \label{sRiem2}\\
^{\left(  3\right)  }R_{\theta\phi\theta\phi}  &  =-\left(  \sin^{2}%
\theta\right)  K_{\theta\theta}^{2} \label{sRiem3}%
\end{align}
It is interesting to note that the component $R_{\theta\phi\theta\phi}$ is
non-null unless the observers are not accelerated $\left(  a=0\right)  $. This
means that the simultaneity hypersurfaces adapted to accelerated observers are
curved regardless to the form of the function $a\left(  \rho\right)  $.

Concerning the cylindrical symmetry, the induced metric is given by%
\begin{equation}
dl^{2}=\left[  \frac{a^{\prime}}{a^{2}}-\left(  1+\frac{a^{\prime}}{a^{2}%
}\right)  \cosh\left(  \eta\right)  \right]  ^{2}d\rho^{2}+G_{\varphi}%
^{2}\left(  \rho\right)  d\phi^{2}+dz^{2}%
\end{equation}
and the extrinsic curvature has the following non-null components:%
\begin{align}
K_{\rho\rho}  &  =\left(  \frac{a^{\prime}}{a}+a\right)  \left[
\frac{a^{\prime}}{a^{2}}-\left(  1+\frac{a^{\prime}}{a^{2}}\right)  \cosh
\eta\right]  \sinh\left(  \eta\right) \\
K_{\phi\phi}  &  =\left[  \frac{1}{a}\cosh\eta-\frac{1}{a}+\rho\right]
\sinh\eta
\end{align}
Now, the only non-null component of the Riemann tensor are $^{\left(
3\right)  }R_{\rho\theta\rho\theta}=-K_{\rho\rho}K_{\phi\phi}$ and the
algebraically equivalent components. Note that, besides the case of null
acceleration, $R_{\rho\theta\rho\theta}$ is zero for a congruence whose
acceleration is $a=\frac{1}{\rho}$. This means that the Rindler congruence is
the unique non-inertial frame in which the adapted simultaneity hypersurfaces
are flat.

For the plane symmetry, we have%
\begin{equation}
dl^{2}=\left[  \frac{a^{\prime}}{a^{2}}-\left(  1+\frac{a^{\prime}}{a^{2}%
}\right)  \cosh\left(  \eta\right)  \right]  ^{2}d\rho^{2}+dy^{2}+dz^{2}%
\end{equation}
and the hypersurfaces are flat for any function $a\left(  \rho\right)  $.

\section{Static observers in curved spacetime}

Intuitively it is expected that, with respect to static observers,
free-falling bodies seem to be accelerated. Evoking a Newtonian picture, this
acceleration can be interpreted as the effect of a local 'gravitational
field'\cite{abramo,sonego}. Thus, motivated by the equivalence principle, we
are led to conjecture that accelerated observers in Minkowski spacetime with
the same proper acceleration of the static observers could simulate aspects of
the gravitational field in their non-inertial frame. This establishes a
connection between static observes of curved spaces and accelerated observers
in Minkowski spacetime. Here we want to examine this connection in detail
considering the Schwarzschild spacetime.

Let us start our discussion recalling some important definitions. A spacetime
is static if it admits a timelike Killing field $T^{\mu}$, which is also
hypersurface-orthogonal \cite{carroll}. In a spacetime that possess this
symmetry, a static observer can be defined as a particle which follows a
worldline whose proper velocity $U^{\mu}=\frac{dx}{d\tau}^{\mu}$ is
proportional to the Killing field:%
\begin{equation}
U^{\mu}=\frac{T^{\mu}}{V} \label{staticO}%
\end{equation}
where $V\left(  x\right)  $ is the normalization function which satisfies
$V^{2}=-T^{\mu}T_{\mu}$.

The reason for this definition is very clear if we consider coordinates
adapted to this Killing field. It can be shown that there exist coordinates in
which the metric can be put in the following form \cite{carroll}:%
\begin{equation}
ds^{2}=g_{00}dt^{2}+g_{ij}dx^{i}dx^{j}%
\end{equation}
where the components $g_{00}$ and $g_{ij}$ do not depend on $t$. In this
coordinates, the Killing field assumes the following simple form: $T^{\mu
}=\left(  1,0,0,0\right)  $. Hence, static observers, according to
(\ref{staticO}), will be characterized by worldlines along which the spatial
coordinates do not change. Note also that $g_{00}=-V^{2}$, in these coordinates.

In a curved spacetime static observers do not follow geodesics. Obviously they
must be accelerated against the 'gravitational attraction' in order to keep
their spatial position unchanged. By using the definition (\ref{staticO}) and
the Killing equation, we can show that the observer's acceleration, $a^{\mu
}=U^{\alpha}\nabla_{\alpha}U^{\mu}$, can be written as \cite{carroll}:
\begin{equation}
a_{\mu}=\nabla_{\mu}\ln V \label{a}%
\end{equation}

Usually the norm of the proper acceleration of an observer is interpreted as
the acceleration that is measured by an instantaneous co-moving free-falling
frame. Indeed, if we denote the quantities in the free-falling frame by a hat
sign and $\Pi^{\mu\nu}$ as the projection operator orthogonal to $U^{\mu}$
(see equation (\ref{pi})) then, the proper acceleration norm can written as
\begin{equation}
a^{2}=g^{\mu\nu}a_{\mu}a_{\nu}=\hat{g}^{\mu\nu}\hat{a}_{\mu}\hat{a}_{\nu
}=\left(  \hat{\Pi}^{\mu\nu}-\hat{U}^{\mu}\hat{U}^{\nu}\right)  \hat{a}_{\mu
}\hat{a}_{\nu}=\hat{\Pi}^{\mu\nu}\hat{a}_{\mu}\hat{a}_{\nu}=\hat{a}_{1}%
^{2}+\hat{a}_{2}^{2}+\hat{a}_{3}^{2}%
\end{equation}
note that we have used the orthogonality condition $U^{\mu}a_{\mu}=0$ and the
fact that in the free-falling frame, the projection tensor assume locally the
diagonal form $\hat{\Pi}^{\mu\nu}=diag(0,1,1,1).$

We should emphasize that this description so far developed here is valid for
all static spacetimes. Now let us concentrate our discussion on the special
case of the Schwarzschild spacetime. Considering the Schwarzschild metric%
\begin{equation}
ds^{2}=-\left(  1-\frac{2GM}{r}\right)  dt^{2}+\left(  1-\frac{2GM}{r}\right)
^{-1}dr^{2}+r^{2}d\Omega^{2},
\end{equation}
where $d\Omega^{2}=d\theta^{2}+\sin^{2}\theta d\phi^{2}$ , it is easy to
verify, by using (\ref{a}), that the proper acceleration field depends on the
position of the static observers according to the formula:
\begin{equation}
a=\frac{GM}{r^{2}}\left(  1-\frac{2GM}{r}\right)  ^{-1/2} \label{aSchw}%
\end{equation}

Here the acceleration is explicitly written in terms of the coordinate $r.$ In
order to make a comparison with the Rindler congruence, it is convenient to
find $a$ as a function of $\rho$ - the distance of the observer's position to
the event horizon located at Schwarzschild radius ($R_{s}=2GM).$ Integrating
the line element in the radial direction from $R_{s}$ to $r$, we find%
\begin{equation}
\rho=r\left(  1-\frac{R_{s}}{r}\right)  ^{1/2}+\frac{1}{2}R_{s}\ln\left[
\frac{2r}{R_{s}}-1+\frac{2r}{R_{s}}\left(  1-\frac{R_{s}}{r}\right)
^{1/2}\right]  \label{rhor}%
\end{equation}
It would be very convenient to write $r$ in terms $\rho$, but the
inverse function cannot be found in an exact form. However,
employing the perturbation method, approximate expressions can be
obtained. For instance, in a domain
close to the horizon, the following expansion is valid%
\begin{equation}
r=R_{s}+\frac{\delta^{2}\left(  \rho\right)  }{R_{s}} \label{rdelta}%
\end{equation}
where $\delta\left(  \rho\right)  $ is some function of $\rho$ which is small
compared to $R_{s}$. Substituting (\ref{rdelta}) in the equation (\ref{rhor})
we obtain, up to order $\delta^{3},$ the result $\rho=2\delta\left(
\rho\right)  +\frac{1}{3R_{s}^{2}}\delta^{3}\left(  \rho\right)  $. Now
considering the expansion of the function $\delta\left(  \rho\right)  $ in
power series, this last equation gives%
\begin{equation}
\delta\left(  \rho\right)  =\frac{1}{2}\rho-\frac{1}{48R_{s}^{2}}\rho^{3}%
\end{equation}
which, combined with equation (\ref{aSchw}) and (\ref{rdelta}) yields the
proper acceleration of static observers that lie in the vicinity the horizon%
\begin{equation}
a=\frac{1}{\rho}-\frac{1}{3R_{s}^{2}}\rho^{3}%
\end{equation}
\qquad\ This result clearly shows that the dependence $1/\rho,$ which holds
for Rindler observers, is valid only approximately for those observers which
are very close to the horizon.

In the opposite case, $r>>R_{s}$, the expression (\ref{rhor}) gives us, in the
first order approximation, $\rho=r$ and, then, equation (\ref{aSchw}) now
yields
\begin{equation}
a=\frac{GM}{\rho^{2}}%
\end{equation}
which, obviously, reproduces the inverse square law predicted by Newtonian
theory of gravity.

The equation (\ref{aSchw}) gives us the proper acceleration that the static
observers in Schwarzschild spacetime are submitted to. In the context of the
present of discussion, we are now led to consider a congruence of observers in
Minkowski spacetime whose acceleration field $a\left(  \rho\right)  $ obeys
the same law (\ref{aSchw}).

The general scheme was already developed in the previous section. The analysis
relative to the Schwarzschild static observers can be promptly done as a
particular case, just by using the appropriate acceleration law (\ref{aSchw}).
Below we show explicitly the embedding functions in terms of the coordinate
$r$:%
\begin{align}
F_{\varphi}\left(  r\right)   &  =2\left(  1-\frac{R_{s}}{r}\right)
^{1/2}\left[  \frac{f\left(  \varphi\right)  }{1-\left(  f\left(
\varphi\right)  \frac{1}{2}\frac{R_{s}}{r^{2}}\right)  ^{2}}\right] \\
G_{\varphi}\left(  r\right)   &  =\left(  1-\frac{R_{s}}{r}\right)
^{1/2}\frac{R_{s}}{r^{2}}\left[  \frac{f^{2}\left(  \varphi\right)
}{1-\left(  f\left(  \varphi\right)  \frac{1}{2}\frac{R_{s}}{r^{2}}\right)
^{2}}\right] \nonumber\\
&  +r\left(  1-\frac{R_{s}}{r}\right)  ^{1/2}+\frac{1}{2}R_{s}\ln\left[
\frac{2r}{R_{s}}-1+\frac{2r}{R_{s}}\left(  1-\frac{R_{s}}{r}\right)
^{1/2}\right]
\end{align}

In Fig.(4), we have plotted on the Minkowski spacetime diagram several
simultaneity sections related to this accelerated congruence. It is
interesting to note, comparing with Fig.(1), that for short distances the
behavior is very similar to the Rindler congruence while that asymptotically
the inverse square law is recovered (see Fig.(2)).

As we have already seen in the previous section, once the embedding functions
are determined, the induced geometry can be directly analyzed by the
techniques presented in the section (2). Considering the spherical symmetry of
Schwarzschild spacetime, equations (\ref{st})-(\ref{sz}) are the natural
choice for the embedding functions. This means that the induced metric is
given by (\ref{dls}) and the curvature of the simultaneity hypersurfaces are
described by the Riemann tensor calculated in (\ref{sRiem1}), (\ref{sRiem2})
and (\ref{sRiem3}), taking into account the extrinsic curvature which is
evaluated in equations (\ref{sK1}), (\ref{sK2}) and (\ref{sK3}).

\section{Field equation for the proper acceleration of static observers}

The connection between statics observers and accelerated frames in
Minkowski spacetime depends crucially on the acceleration of the
static observes. In this section we want to deduce the field
equation that dictates the behavior of proper acceleration field
in curved spaces. This can be achieved by rewriting Einstein
equations conveniently. With this purpose, we are going to use the
embedding formalism once again. But now we should keep in mind
that the ambient space $M$ is a static curved manifold and the
hypersurface are the 3-dimensional space orthogonal to the
timelike Killing field $T^{\mu}$.

An important characteristic of this embedding is that the extrinsic curvature
is null. In order to check this, consider the covariant derivative of $U^{\mu
}:$
\begin{equation}
\nabla_{\mu}U_{\nu}=\frac{1}{V}\left(  \nabla_{\mu}T_{\nu}-T_{\nu}\nabla_{\mu
}\ln V\right)
\end{equation}

It happens that, for a hypersurface-orthogonal Killing field, we may write
\cite{wald}%
\begin{equation}
\nabla_{\mu}T_{\nu}=T_{\nu}\nabla_{\mu}\ln V-T_{\mu}\nabla_{\nu}\ln V
\end{equation}
It follows immediately that%
\begin{equation}
\nabla_{\mu}U_{\nu}=-U_{\mu}a_{\nu} \label{DU}%
\end{equation}
which, according to the definition (\ref{K}), leads to $K_{ab}=0$.

As a consequence, the Gauss-Codazzi equation reduces to%
\begin{align}
^{\left(  3\right)  }R_{abcd}  &  =e_{a}^{\alpha}e_{b}^{\beta}e_{c}^{\gamma
}e_{d}^{\delta}R_{\alpha\beta\gamma\delta}\\
U^{\mu}e_{a}^{\alpha}e_{b}^{\beta}e_{c}^{\gamma}R_{\mu\alpha\beta\gamma}  &
=0
\end{align}

Now in order to obtain the Einstein equations, let us first determine the
Ricci tensor. Here we are going to follow closely the notation of reference
\cite{seahra}. Contracting the first and third indices of the Riemann tensor
using the spacetime metric conveniently written as $g^{\alpha\gamma}%
=e_{a}^{\alpha}e_{c}^{\gamma}h^{ac}-U^{\alpha}U^{\gamma}$, we obtain
\begin{equation}
R_{\beta\nu}=\left(  h^{ac}e_{a}^{\alpha}e_{c}^{\gamma}R_{\alpha\beta\gamma
\nu}-U^{\alpha}U^{\gamma}R_{\alpha\beta\gamma\nu}\right)
\end{equation}
Each index of the Ricci tensor can be projected either in the
tangential direction of $\Sigma,$ given by $e_{b}^{\beta}$ , or
along the orthogonal direction $U^{\mu}$. Taking into account the
Gauss-Codazzi equations, these
projections yield \cite{seahra}%
\begin{align}
e_{i}^{\mu}e_{j}^{\nu}R_{\beta\mu}  &  =\,^{\left(  3\right)  }R_{ij}%
-\,^{\left(  3\right)  }E_{ij}\label{Rij}\\
R_{\mu\nu}U^{\mu}e_{j}^{\nu}  &  =0\label{RjU}\\
R_{\mu\nu}U^{\mu}U^{\nu}  &  =\,^{\left(  3\right)  }E_{ij}h^{ij} \label{RUU}%
\end{align}
where $^{\left(  3\right)  }R_{ij}$ is the intrinsic Ricci tensor of $\Sigma$
and $^{\left(  3\right)  }E_{ij}$ is a symmetric tensor defined as%
\begin{equation}
^{\left(  3\right)  }E_{ij}=U^{\mu}U^{\nu}e_{i}^{\alpha}e_{j}^{\beta}%
R_{\alpha\mu\beta\nu} \label{E}%
\end{equation}
This tensor inhabits the hypersurface $\Sigma$, but it is not an intrinsic
geometric quantity of the hypersurface since it depends on the orthogonal
components of Riemann tensor defined in the ambient space $M$. As we shall
see, $^{\left(  3\right)  }E_{ij}$ can be written in terms of the components
of the proper acceleration. Indeed, from the definition of the Riemann tensor,
we have%
\begin{equation}
R_{\alpha\mu\beta\nu}U^{\mu}=D_{\beta}D_{\nu}U_{\alpha}-D_{\nu}D_{\beta
}U_{\alpha}%
\end{equation}
By using (\ref{DU}) in the above equation and contracting it with $U^{\nu
}e_{a}^{\alpha}e_{b}^{\beta}$, we find%
\begin{equation}
^{\left(  3\right)  }E_{ij}=a_{i}a_{j}+e_{i}^{\mu}e_{j}^{\nu}D_{\mu}a_{\nu}%
\end{equation}
where, for the sake of simplicity, we are using the notation: $a_{i}%
=e_{i}^{\alpha}a_{\alpha}.$

The acceleration vector $a^{\mu}$ is orthogonal to the proper velocity
$U^{\mu}$, this means that it belongs to the tangent space of $\Sigma$. Thus,
by using the definition of induced covariant derivative (see equation
(\ref{CovD})), it possible to write
\begin{equation}
^{\left(  3\right)  }E_{ij}=a_{i}a_{j}+\,^{\left(  3\right)  }\nabla_{i}a_{j}%
\end{equation}

Thus, the equations (\ref{Rij}) and (\ref{RUU}) assume the form%
\begin{align}
e_{i}^{\mu}e_{j}^{\nu}R_{\mu\nu}  &  =\,^{(3)}R_{ij}-a_{i}a_{j}-\,^{\left(
3\right)  }\nabla_{i}a_{j}\label{eqRij}\\
U^{\mu}U^{\nu}R_{\mu\nu}  &  =a^{2}+\,^{\left(  3\right)  }\nabla_{i}a^{i}
\label{eqa}%
\end{align}

Taking into account the Einstein equations, $R_{\mu\nu}=8\pi G\left(
T_{\mu\nu}-\frac{1}{2}g_{\mu\nu}T\right)  $, we can say that the above
equations represent the field equations for the proper acceleration of static
observers. In particular the equation (\ref{eqa}) has a very interesting form
that resembles the Newtonian gravitational field equation. However we have to
highlight three important modifications due to the relativistic effects: i)The
energy density $\left(  T_{\mu\nu}U^{\mu}U^{\nu}\right)  $ is no longer the
exclusive source of gravity, since there is now contributions that come from
other components of the energy-momentum tensor; ii)There is the non-linear
term $a^{2}$, whose presence in (\ref{eqa}) demonstrates that gravity is a
self-interacting field in the relativistic regime; and iii) there is the fact
that the divergence of $a^{i}$ depends on the intrinsic geometry of the
hypersurface $\Sigma$, which can be curved.

We should also point out that equation (\ref{RjU}) does not constrain neither
the acceleration nor the metric of $\Sigma$, but, on the other hand, it
imposes the natural condition according to which there is no energy flux in
static spacetimes as seen from static observers.

Now, in order to proceed further, it is important to find a relation between
the scalar curvature of $M$ and the intrinsic scalar curvature of $\Sigma.$
With the help of equations (\ref{Rij}),(\ref{RjU}),(\ref{RUU}), it can be
shown that
\begin{equation}
R=^{(3)}R-2\left(  a^{2}+\nabla_{i}a^{i}\right)  \label{R}%
\end{equation}

Now, equations (\ref{Rij}),(\ref{RjU}),(\ref{RUU}) and (\ref{R}) permit us to
examine the decomposition of the Einstein tensor. It is not difficult to check
that the components of the Einstein tensor can be written as%
\begin{align}
G_{\mu\nu}U^{\mu}U^{\nu}  &  =\frac{1}{2}^{(3)}R\label{GUU}\\
G_{\mu\nu}U^{\mu}e_{i}^{\nu}  &  =0\label{GiU}\\
G_{ij}  &  =^{(3)}G_{ij}-a_{i}a_{j}-\nabla_{i}a_{j}+h_{ij}\left(  a^{2}%
+\nabla_{i}a^{i}\right)  \label{Gij}%
\end{align}
where we are using again the notation $G_{ij}=e_{i}^{\alpha}e_{j}^{\beta
}G_{\alpha\beta}$.

In the previous sections, we deal with acceleration fields described by
functions of a unique coordinate, by virtue of the symmetries of the
congruence. Thus, in order to make contact with the previous discussion, let
us admit that the static spacetime exhibit additional symmetries.

So suppose that $M$ possesses a spacelike Killing vector field $X^{\mu}$
orthogonal to the vector field $T^{\mu}$. Furthermore, let us admit that it
commutates with the time-translation generator, $T^{\mu}$. Of course, this
means that $X^{\mu}$ satisfies the equation:
\begin{equation}
T^{\alpha}\nabla_{\alpha}X_{\mu}=X^{\alpha}\nabla_{\alpha}T_{\mu}%
\end{equation}

We shall show that $X^{\mu}$ is also orthogonal to $a^{\mu}$. Indeed, taking
the inner product with $T^{\mu}$ and using the Killing equation for $X^{\mu}$,
the above equation gives $X^{\alpha}\nabla_{\alpha}V^{2}=0$ , which, recalling
the equation (\ref{a}), implies the orthogonality condition:%
\begin{equation}
X^{\alpha}a_{\alpha}=0 \label{Xa}%
\end{equation}

As consequence, we shall see that $X^{\mu}$ is a vector that belongs to the
tangent space of the level surfaces of the function $V\left(  x\right)  $.
Indeed, in any one of the hypersurfaces of $M$ that are orthogonal to the
timelike Killing vector field $T^{\mu}$, (remember that these hypersurfaces
are isometric 3-manifolds, since they do not evolve in a static spacetime),
the equation $V\left(  x\right)  =c$, where $c$ is some positive constant,
defines a level surface of the function $V\left(  x\right)  $, which we denote
by $S_{c}$. Of course equation (\ref{a}) implies that $a^{\mu}$ is
perpendicular to this level surface. Thus, considering that $X^{\mu}$ is also
orthogonal to $T^{\mu}$, the equation (\ref{Xa}) leads us to conclude that
$X^{\mu}$ lies in the tangent space of $S_{c}$ for some $c$. This holds in
every point of $M$.

In general, the norm of acceleration vector $a^{\mu}$ may vary in the surface
$S_{c}.$ However, it is easy to show that, in the direction of Killing field
$X^{\mu},$ the derivative of the norm $a=\sqrt{a_{\mu}a^{\mu}}$ is null:%
\begin{equation}
X^{\mu}\nabla_{\mu}a=0
\end{equation}

If there exist at least two linearly independent Killing field like this
vector $X^{\mu}$, then, as they will span the two-dimensional tangent space of
$S_{c}$, it follows that derivative of $a$ is zero along any tangent direction
of $S_{c}$. In other words, this means that $S_{c}$ is also a level surface
for the acceleration $a.$ Therefore, we can write%
\begin{equation}
a=Y\left(  V\right)  \label{afV}%
\end{equation}
for some function $Y$.

Now consider the normal vector of $S_{c}$, i.e, $\sigma^{\mu}=\frac{a^{\mu}%
}{a}.$ Since $S_{c}$ is a level surface of $a$, the gradient of $a$ is
co-linear to $\sigma^{\mu}$. This fact allows us to write%
\begin{equation}
\nabla_{\mu}a=\left(  \sigma^{\nu}\nabla_{\nu}a\right)  \sigma_{\mu}%
\end{equation}
Based on this and also on the condition $\nabla_{\mu}a_{\nu}=\nabla_{\nu
}a_{\mu}$ (that follows directly from equation (\ref{a})), we can verify that
$\sigma^{\mu}$ satisfies the geodesic equation%
\begin{equation}
\sigma^{\mu}\nabla_{\mu}\sigma^{\nu}=0 \label{geo}%
\end{equation}
As the embedding of $\Sigma$ into $M$ has null extrinsic curvature then the
above equation is equivalent to $\sigma^{i}\left(  ^{3}\nabla_{i}\sigma
^{j}\right)  =0$, i.e., the vector $\sigma^{i}=e_{\mu}^{i}\sigma^{\mu}$
satisfies the geodesic equation with respect to the induced geometry of the
hypersurface $\Sigma.$

Let us now explore the consequences of this condition. Pick a certain value
for $c.$ As we know, associated to this value corresponds a particular level
surface $S_{c}$. Let $x_{\bot}^{\left(  1\right)  }$ and $x_{\bot}^{\left(
2\right)  }$ be intrinsic coordinates of $S_{c}$. Solving the equation
(\ref{geo}) for geodesics that cross $S_{c}$ perpendicularly in every point,
we can use the affine parameter of these curves, $\rho$, together with
coordinates$\left\{  x_{\bot}\right\}  $, to build a Gaussian normal
coordinate system adapted to $S_{c}$ in its neighborhood. It follows that, in
these new coordinate system, $V$ depends exclusively on $\rho$.

To check this, first note that, in these coordinates, $\sigma^{\mu}=\left(
\frac{\partial x^{\mu}}{\partial\rho}\right)  _{x_{\bot}}$. Now, from the
normalization condition for $\sigma^{\mu}$, conveniently written as
\begin{equation}
\frac{a_{\mu}}{a}\left(  \frac{\partial x}{\partial\rho}^{\mu}\right)
_{x_{\bot}}=1
\end{equation}
and recalling equation (\ref{a}) and (\ref{afV}), we find the equation%
\begin{equation}
\frac{1}{Y\left(  V\right)  }\left(  \frac{\partial\ln V}{\partial\rho
}\right)  _{x_{\bot}}=1 \label{eqVrho}%
\end{equation}
If we assume, as initial condition, that $V$ is a uniform function
in that chosen surface $S_{c},$, then, it follows, from the
solution of equation (\ref{eqVrho}), that, in some neighborhood of
$S_{c}$, $V$ is a function of coordinate $\rho$ only. Thus, we can
write $V=V\left(  \rho\right)  $.

Therefore the spacetime metric, written in these coordinates, assume the
following simple form%
\begin{equation}
ds^{2}=V^{2}\left(  \rho\right)  dt^{2}+d\rho^{2}+\gamma_{AB}dx_{\bot}%
^{A}dx_{\bot}^{B} \label{metricrho}%
\end{equation}
where the capital indices have the following range $A,B$ =1,2. We should note
that, depending on the commutation relation between the Killing fields that
lies in $S_{c}$, the induced metric $\gamma_{AB}$ in the level surfaces may
admit extra simplifications. For instance, Schwarzschild metric is a special
case of (\ref{metricrho}) in which level surfaces has spherical symmetry.

Now considering the metric in this special form (\ref{metricrho}), we are led
to examine the question of classifying the possible sources of this kind of
spacetime. In other words, we have to faced the task of characterizing the
energy-momentum of matter distributions that could generate such metrics. In
particular we are interested in determining the conditions the energy-momentum
tensor should satisfy in order to static observers in a curved space have the
same acceleration, $a\left(  \rho\right)  =1/\rho,$ as the Rindler observers
in Minkowski spacetime.

The better approach to deal with this question is to express $^{\left(
3\right)  }G_{ij}$ and $^{\left(  3\right)  }R,$ that appears in the equations
(\ref{GUU}), (\ref{GiU}) and (\ref{Gij}), in terms of intrinsic and extrinsic
curvatures of the level surfaces $S_{c}$, applying the embedding formalism
again. Making some minor adjustment, the same procedure previously described
can be repeated and now it yields the following equations\cite{mcmanus}:
\begin{align}
^{\left(  3\right)  }G_{ij}\sigma^{i}\sigma^{j}  &  =-\frac{1}{2}\,^{\left(
2\right)  }R-\frac{1}{2}\Omega^{AC}\Omega_{AC}+\frac{1}{2}\Omega^{2}\\
^{(3)}G_{ij}L_{A}^{i}\sigma^{j}  &  =\,^{(2)}\nabla_{C}\Omega_{A}%
^{C}-\,^{\left(  2\right)  }\nabla_{A}\Omega\\
^{\left(  3\right)  }G_{ij}L_{A}^{i}L_{B}^{j}  &  =\,^{\left(  2\right)
}G_{AB}-\sigma^{c}\nabla_{c}\left(  \Omega_{AB}-\gamma_{AB}\Omega\right)
+\nonumber\\
&  +\left(  2\Omega_{A}^{C}\Omega_{BC}-3\Omega\Omega_{AB}\right)  +\frac{1}%
{2}\gamma_{AB}\left(  \Omega^{CD}\Omega_{CD}+\Omega^{2}\right) \nonumber\\
^{\left(  3\right)  }R  &  =\,^{\left(  2\right)  }R-\left(  2\sigma^{c}%
\nabla_{c}\Omega+\Omega_{AB}\Omega^{AB}+\Omega^{2}\right)
\end{align}
where $\sigma^{i}$ is the spacelike normal vector of the level
surfaces $S_{c}$, $^{\left(  2\right)  }\nabla$ is the induced
covariant derivative compatible with the induced metric
$\gamma_{AB}$ of $S_{c}$, $^{\left( 2\right)  }G_{AB}$ and
$^{\left(  2\right)  }R$ are the intrinsic Einstein tensor and
scalar curvature of $S_{c}$ surfaces respectively, the symmetric
tensor $\Omega_{AB}$ is the extrinsic curvature of the level
surfaces embedded in $\Sigma$ and finally $L_{A}^{i}$ is the
analogue of $e_{a}^{\mu}$ (see equation (\ref{e})) relative to the
embedding of the level surfaces $S_{c}$ into the hypersurface
$\Sigma$.

It is important to emphasize here that, in order to deduce these relations, we
had to handle the tensor $^{\left(  2\right)  }E_{AB}$ whose definition%
\begin{equation}
^{\left(  2\right)  }E_{BD}=\,^{\left(  3\right)
}R_{abcd}\sigma^{a}\sigma
^{c}L_{B}^{b}L_{D}^{d}%
\end{equation}
is the two-dimensional analogue of (\ref{E}). By using the fact that
$\sigma^{i}$ satisfies the geodesic equation, we could show, from the
definition of the Riemann tensor, that%
\begin{equation}
^{\left(  2\right)  }E_{BD}=-\sigma^{c}\nabla_{c}\Omega_{BD}+\Omega_{B}%
^{C}\Omega_{CD} \label{E2}%
\end{equation}
Another useful relation we have employed too was the following formula (valid
in the Gaussian coordinate system):%
\begin{equation}
\Omega_{AB}=\frac{1}{2}\sigma^{c}\nabla_{c}\gamma_{AB}%
\end{equation}

Now, before we introduce the energy-momentum tensor into the Einstein
equations, let us make one additional assumption concerning the geometry of
the spacetime $M$. For the sake of simplicity, henceforth we shall admit that
the level surfaces $S_{c}$ are maximally symmetric spaces. As a consequence,
the Riemann tensor and the extrinsic curvature reduce to these simple forms%
\begin{align}
^{\left(  2\right)  }R_{ABCD}  &  =\frac{^{(2)}R}{2}\left(  \gamma_{AC}%
\gamma_{BD}-\gamma_{AD}\gamma_{BC}\right) \\
\Omega_{AB}  &  =\frac{\Omega}{2}\gamma_{AB}%
\end{align}
where $^{(2)}R$ and $\Omega$ (trace of extrinsic curvature) do not depend on
the intrinsic coordinates of $S_{c},$ but may depend on $\rho$. The value of
$\,^{\left(  2\right)  }R$ determines completely the intrinsic geometry of the
level surfaces. It is well known that if $\,^{\left(  3\right)  }R$ is
positive, null or negative then, the geometry of $S_{c}$ is spherical, plane
or hyperbolic, respectively.

A direct consequence of this assumption is that the Einstein tensor can be
totally written as a function of the proper acceleration, the scalar curvature
and extrinsic curvature of $S_{c}$ surfaces in the following form:
\begin{align}
G_{\mu\nu}U^{\mu}U^{\nu}  &  =\frac{1}{2}\,^{\left(  2\right)  }R-\frac{3}%
{4}\Omega^{2}-\Omega^{\prime}\label{EUU}\\
G_{\mu\nu}U^{\mu}e_{i}^{\nu}  &  =0\label{EUi}\\
G_{ij}\sigma^{i}\sigma^{j}  &  =-\frac{1}{2}\,^{\left(  2\right)  }R+\frac{1}%
{4}\Omega^{2}+a\Omega\label{Eij}\\
G_{ij}\sigma^{j}L_{A}^{i}  &  =0\label{EiA}\\
G_{ij}L_{A}^{i}L_{B}^{j}  &  =\left[  \frac{1}{2}\Omega^{\prime}+\frac{1}%
{4}\Omega^{2}+\left(  a^{2}+a^{\prime}\right)  +\frac{1}{2}a\Omega\right]
\gamma_{AB} \label{EAB}%
\end{align}
where the prime denote derivative with respect to $\rho$.

We should also note that, due to the Bianchi identities, the scalar curvature
and the extrinsic curvature of the maximally symmetric surfaces $S_{c}$ must
satisfy the equation:%
\begin{equation}
^{\left(  2\right)  }R^{\prime}+\,^{\left(  2\right)  }R\Omega=0
\end{equation}

Based on the form of the Einstein tensor, we can conclude that the
most general source of this special class of static spacetimes (in
which the level surfaces of the norm of static observers'
acceleration are maximally symmetric two-dimensional spaces) is a
non-viscous fluid described by the following
energy-momentum tensor%
\begin{equation}
T_{\mu\nu}=\left(  \epsilon+p_{\bot}\right)  U_{\mu}U_{\nu}+\left(  p_{\Vert
}-p_{\bot}\right)  \sigma_{\mu}\sigma_{\nu}+p_{\bot}g_{\mu\nu} \label{T}%
\end{equation}
where $\epsilon$ is the energy density, $p_{\Vert}$ is the pressure in the
parallel direction to acceleration of static observers and $p_{\bot}$ is the
pressure in the orthogonal direction relative to $\sigma^{i}$. All the
quantities are measured by the static observers.

Combining equations (\ref{EUU})-(\ref{EAB}) and (\ref{T}), the Einstein
equations reduces to
\begin{align}
\frac{1}{2}\,^{\left(  2\right)
}R-\frac{3}{4}\Omega^{2}-\Omega^{\prime}  &
=8\pi G\epsilon\label{eq1}\\
-\frac{1}{2}\,^{\left(  2\right)  }R+\frac{1}{4}\Omega^{2}+a\Omega
&  =8\pi
Gp_{\Vert}\label{eq2}\\
\frac{1}{2}\Omega^{\prime}+\frac{1}{4}\Omega^{2}+\left(  a^{2}+a^{\prime
}\right)  +\frac{1}{2}a\Omega &  =8\pi Gp_{\bot} \label{eq3}%
\end{align}

Besides, the fluid must satisfies, as a consequence of the energy-momentum
conservation, the following equation%
\begin{equation}
p_{\Vert}^{\prime}+\left(  \epsilon+p_{\Vert}\right)  a+\left(  p_{\Vert
}-p_{\bot}\right)  \Omega=0 \label{conservation}%
\end{equation}

Once these equations have been established, we are now in an appropriate
position to attack the problem concerning the existence of a curved static
spacetime in which the 'Rindler law' is reproduced exactly. As we have seen,
the inverse law is satisfied in the Schwarzschild spacetime only approximately
by those static observes who are near the horizon. Now, we want to investigate
whether there is any source capable of generating a spacetime where the
acceleration of static observer is precisely given by $a=\frac{1}{\rho}$. It
is easy to see that $a^{2}+a^{\prime}=0$ for a Rindler acceleration and, then,
the equations (\ref{eq1}), (\ref{eq2}) and (\ref{eq3}) allow us to write the
extrinsic and intrinsic curvature directly in terms of the energy density and
the pressures of the fluid as
\begin{align}
\Omega &  =4\pi G\rho\left(  \epsilon+p_{\Vert}+2p_{\bot}\right)
\label{omega}\\
^{\left(  2\right)  }R  &  =8\pi G\left(  \epsilon-p_{\Vert}+2p_{\bot}\right)
+\frac{1}{2}\left[  4\pi G\rho\left(  \epsilon+p_{\Vert}+2p_{\bot}\right)
\right]  ^{2} \label{R2}%
\end{align}
where the energy density and the pressures are not totally independent since
they must satisfy the conservation equation. A simple solution of equation
(\ref{conservation}) is obtained by taking the following state equation%
\begin{align}
p_{\Vert}  &  =-\epsilon_{0}\label{pressureP}\\
p_{\bot}  &  =0 \label{pressureO}%
\end{align}
where $\epsilon_{0}$ is a uniform energy density that should be
positive in order to satisfy the weak energy condition. These
polytropic state equations imply, according to equations
(\ref{omega}) and (\ref{R2}), that the level surfaces have null
extrinsic curvature $\left(  \Omega=0\right)  $ and a positive
intrinsic curvature given by $^{\left(  2\right)  }R=16\pi
G\epsilon_{0}$.

It is interesting to note that these are sufficient conditions to completely
determine the curvature of this particular spacetime $M$. Indeed, by using the
embedding formalism, we can show that the non-null components of the Riemann
tensor are reduced to these following components%
\begin{equation}
R_{ijkl}L_{A}^{i}L_{B}^{j}L_{C}^{k}L_{D}^{l}=8\pi G\epsilon_{0}\left(
\gamma_{AC}\gamma_{BD}-\gamma_{AD}\gamma_{BC}\right)
\end{equation}
where $R_{ijkl}=R_{\mu\nu\alpha\beta}e_{i}^{\mu}e_{j}^{\nu}e_{k}^{\alpha}%
e_{l}^{\beta}$. Therefore, $M$ is flat only in the empty space. If
$\epsilon_{0}$ is non-null, then, $M$ is curved and it is such that static
observers follow the Rindler acceleration law.

Finally, let us make some comments concerning the source of this
special spacetime. As it is well known, the state equation
$p=-\epsilon$ is characteristic of a cosmological term or of an
energy-momentum tensor related to a field that is found in a false
vacuum state. The state equation (\ref{pressureP}) shows a similar
dependence between the parallel pressure and energy density,
however because of (\ref{pressureO}) it is clear that the fluid is
not isotropic. Hence, we may say that the source is a kind of
anisotropic false vacuum state. It is worthy of mention that some
inflationary models speculate about the possibility that the
Universe has passed through such rather exotic phase at the early
stages of the cosmic evolution \cite{jensen}.

\section{Summary}

In curved static spacetime, static observers have a non-zero
proper acceleration. From their perspective, free-falling bodies
are accelerated due to the attraction of a local gravitational
field. Because of this interpretation and motivated by the
equivalence principle, we have studied congruences, defined in
Minkowski spacetime, composed of timelike curves whose
acceleration field coincides to the acceleration field of the
static observers. Our purpose is to simulate in the non-inertial
frame defined in Minkowski spacetime some aspects of the
gravitational field that is experienced by static observers in
curved spaces.

The embedding of the simultaneity hypersurfaces adapted to the non-inertial
frame associated to the congruences, which is an important element in this
context, was determined explicitly for any acceleration field, based on the
locality principle. We have also investigated the induced geometry of the
simultaneity hypersurfaces, determining explicitly the induced metric and the
intrinsic Riemann tensor.

The particular case of the Schwarzschild spacetime was investigated in detail.
The simultaneity sections were determined and illustrated in Fig. (4). This
figure shows clearly that the congruence is equivalent to the Rindler frame
only near the event horizon Fig. (1). While that, at large distances, it
behaves like a congruence whose acceleration field obeys the inverse square
law (see Fig.(2)).

We have also obtained the field equation that regulates the
behavior of accelerated field of static observers from the
Einstein equations. We have shown that this equation can be
understood as the relativistic version of the Newtonian
gravitational field equation. As we have seen, according to this
equation the divergence (calculated with respect to the induced
geometry of the simultaneity hypersurface) of the acceleration
field depends on the effective energy density (energy density plus
the trace of energy-momentum tensor) and on the norm squared of
the acceleration field. The presence of this non-linear term is
clearly a consequence of the self-interacting property of gravity
in the relativistic regime.

When the spacetime is characterized by the fact that level surfaces of the
proper acceleration are maximally symmetric spaces, the field equations are
simplified and can be totally expressed in terms of the intrinsic and
extrinsic curvature of the level surfaces and the acceleration field. The
general form of tensor-energy momentum tensor of the possible sources for this
kind of static spacetime was determined. We have seen that it corresponds to a
non-viscous fluid totally characterized by the energy density, the parallel
pressure and orthogonal pressure, which, in principle, may be different.

When the state equation of the fluid is that of an anisotropic false vacuum
state, the produced spacetime has curvature and it is such that static
observes are accelerated exactly according to the inverse law characteristic
of the Rindler congruence.

\newpage

\section{Appendix}

Here we have a list of figures illustrating the embedding of the
simultaneity hypersurfaces into Minkowski spacetime adapted to
different congruences of
accelerated observers.%

%TCIMACRO{\FRAME{fhFU}{1.9545in}{1.9545in}{0pt}{\Qcb{{\small Non-inertial frame
%associated to the Rindler congruence which satisfies the inverse law }$\left(
%{\small a=}\frac{1}{\rho}\right)  ${\small . The dashed lines are hyperbolas
%that represent the worldlines of some observers. The full lines are
%simultaneity hypersurfaces adapted to the non-inertial frame corresponding to
%different instants of time (}${\small f}\left(  \varphi\right)  {\small =0.1}$
%{\small and }${\small f}\left(  \varphi\right)  {\small =0.3},$
%{\small respectively)}}}{\Qlb{fig1}}{invrho.eps}%
%width 1.9545in;  height 1.9545in;  depth 0pt;  original-width 7.7089in;
%original-height 7.7089in;  cropleft "0";  croptop "1";  cropright "1";
%cropbottom "0";  filename 'invrho.eps';file-properties "XNPEU";}}}%
%BeginExpansion
\begin{figure}
[h!]
\begin{center}
\includegraphics[
height=1.9545in,
width=1.9545in
]%
{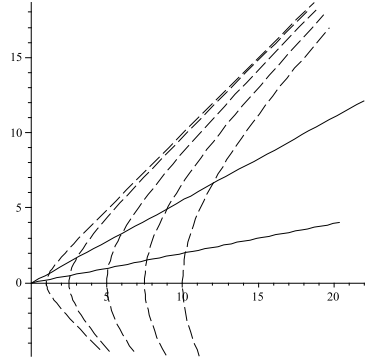}%
\caption{{\small Non-inertial frame associated to the Rindler
congruence which satisfies the inverse law ($ a=\frac{1}{\rho} $)
. The dashed lines are hyperbolas that represent the worldlines of
some observers. The full lines are simultaneity hypersurfaces
adapted to the non-inertial frame corresponding to different
instants of time ( $ f\left(
\varphi\right)=0.1 $ and $ f \left(  \varphi\right)=0.3,$ respectively)}}%
\label{fig1}%
\end{center}
\end{figure}
%EndExpansion
%

%TCIMACRO{\FRAME{fbhFU}{1.9536in}{1.9536in}{0pt}{\Qcb{{\small Non-inertial
%frame associated to a congruence that obey the inverse square law }$\left(
%{\small a=}\frac{1}{\rho^{2}}\right)  ${\small . The dashed lines are
%hyperbolas that represent the wordlines of some observers. The full lines are
%simultaneity hypersurfaces adapted to the non-inertial frame corresponding to
%different instant of time (}${\small f}\left(  \varphi\right)  {\small =0.5}$
%{\small and }${\small f}\left(  \varphi\right)  {\small =1.2},$
%{\small respectively)}}}{\Qlb{fig2}}{invrho2.eps}%
%width 1.9536in;  height 1.9536in;  depth 0pt;  original-width 7.7089in;
%original-height 7.7089in;  cropleft "0";  croptop "0.9993";
%cropright "0.9993";  cropbottom "0";
%filename 'invrho2.eps';file-properties "XNPEU";}}}%
%BeginExpansion
\begin{figure}
[h!]
\begin{center}
\includegraphics[
trim=0.000000in 0.000000in 0.005396in 0.005396in,
height=1.9536in,
width=1.9536in
]%
{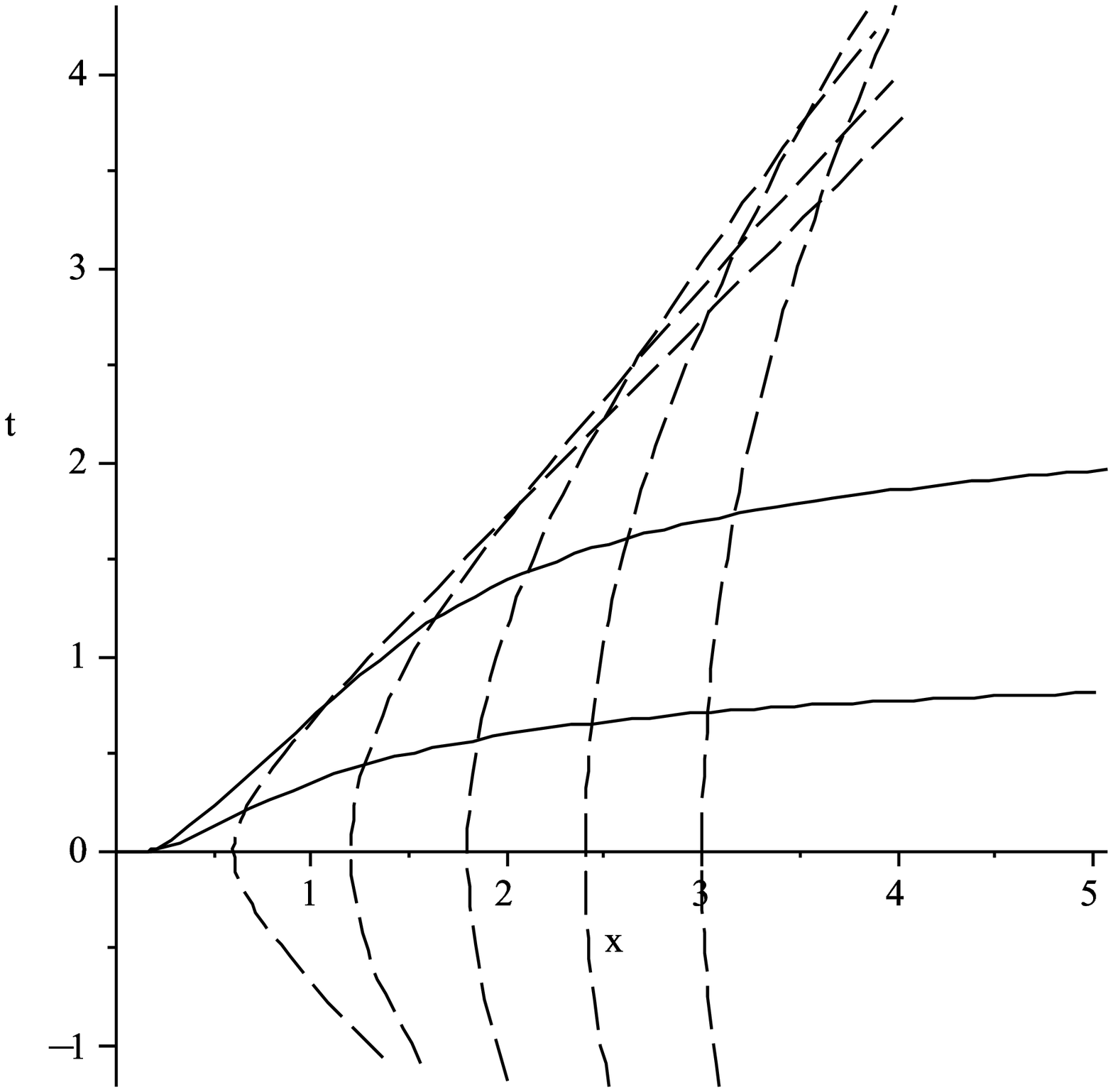}%
\caption{{\small Non-inertial frame associated to a congruence
that obey the inverse square law ($ a=\frac{1}{\rho^{2}} $) . The
dashed lines are hyperbolas that represent the wordlines of some
observers. The full lines are simultaneity hypersurfaces adapted
to the non-inertial frame corresponding to different instants of
time ( $ f\left( \varphi\right)  =0.5 $ and $ f
\left(  \varphi\right)  =1.2$, respectively)}}%
\label{fig2}%
\end{center}
\end{figure}
%EndExpansion

\newpage%

%TCIMACRO{\FRAME{fthFU}{1.9536in}{1.9536in}{0pt}{\Qcb{{\small Non-inertial
%frame associated to a congruence of observers submitted to the same
%acceleration }${\small a=1}${\small . The dashed lines are hyperbolas that
%represent the worldlines of some observers. The full lines are simultaneity
%hypersurfaces adapted to the non-inertial frame corresponding to different
%instants of time (}${\small f}\left(  \varphi\right)  {\small =0.1}%
%${\small and }${\small f}\left(  \varphi\right)  {\small =0.3},$
%{\small respectively)}}}{\Qlb{fig3}}{aconst.eps}%
%width 1.9536in;  height 1.9536in;  depth 0pt;  original-width 7.7089in;
%original-height 7.7089in;  cropleft "0";  croptop "0.9993";
%cropright "0.9993";  cropbottom "0";
%filename 'aconst.eps';file-properties "XNPEU";}}}%
%BeginExpansion
\begin{figure}
[h!]
\begin{center}
\includegraphics[
trim=0.000000in 0.000000in 0.005396in 0.005396in,
height=1.9536in,
width=1.9536in
]%
{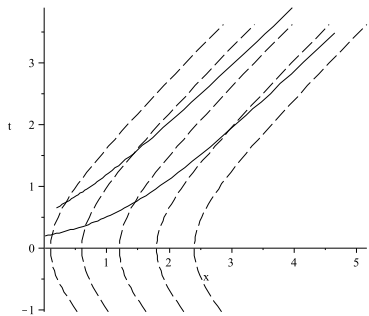}%
\caption{{\small Non-inertial frame associated to a congruence of
observers submitted to the same acceleration $ a=1 $ . The dashed
lines are hyperbolas that represent the worldlines of some
observers. The full lines are simultaneity hypersurfaces adapted
to the non-inertial frame corresponding to different instants of
time ($ f\left(  \varphi\right) =0.1 $ and $ f\left(
\varphi\right)  =0.3 $,
respectively)}}%
\label{fig3}%
\end{center}
\end{figure}
%EndExpansion
%

%TCIMACRO{\FRAME{fbhFU}{1.9545in}{1.9545in}{0pt}{\Qcb{{\small Non-inertial
%frame associated to a congruence of timelike curves whose acceleration field
%coincides with the acceleration of static observers in the Schwarzschild
%spacetime }$\left(  {\small a=}\frac{GM}{r^{2}}\left(  1-\frac{2GM}{r}\right)
%^{-1/2}\right)  ${\small . The dashed lines are hyperbolas that represent the
%worldlines of some observers. We have assumed }${\small GM=1}${\small . The
%full lines are simultaneity hypersurfaces adapted to the non-inertial frame
%corresponding to different instants of time (}${\small f}\left(
%\varphi\right)  {\small =1}$ {\small and }${\small f}\left(  \varphi\right)
%{\small =2.8}${\small ,} {\small respectively). Note the equivalence with the
%Rindler congruence, for small }${\small \rho}${\small . At large distances,
%the behavior is that of a congruence whose acceleration field obey the inverse
%square law.}}}{\Qlb{fig4}}{aschw.eps}{\special{ language "Scientific Word";
%type "GRAPHIC";  maintain-aspect-ratio TRUE;  display "USEDEF";
%valid_file "F";  width 1.9545in;  height 1.9545in;  depth 0pt;
%original-width 7.7089in;  original-height 7.7089in;  cropleft "0";
%filename 'aSchw.eps';file-properties "XNPEU";}}}%
%BeginExpansion
\begin{figure}
[h!]
\begin{center}
\includegraphics[
height=1.9545in,
width=1.9545in
]%
{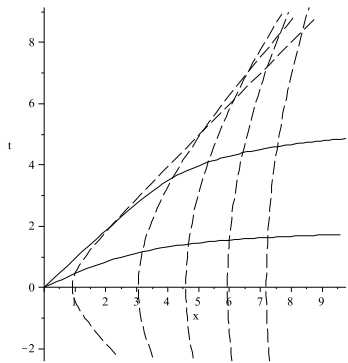}%
\caption{{\small Non-inertial frame associated to a congruence of timelike
curves whose acceleration field coincides with the acceleration of static
observers in the Schwarzschild spacetime ($ a=\frac{GM}{r^{2}%
}\left(  1-\frac{2GM}{r}\right)  ^{-1/2}  $). The dashed lines are
hyperbolas that represent the worldlines of some observers. We
have assumed $ GM=1 $. The full lines are simultaneity
hypersurfaces adapted to the non-inertial frame corresponding to
different instants of time ( $f \left( \varphi\right)  =1 $ and $
f \left( \varphi\right) =2.8 $ , respectively). Note the
equivalence with the Rindler congruence, for small $ \rho $. At
large distances, the behavior is that of a
congruence whose acceleration field obeys the inverse square law.}}%
\label{fig4}%
\end{center}
\end{figure}
%EndExpansion

\newpage

\section{Acknowledgement}

The authors thank CNPq for financial support.

\end{document}